\documentclass[12pt]{article}
\pdfoutput=1
\usepackage{float}
\usepackage{authblk}
\usepackage{amssymb,amsmath}
\usepackage{graphicx}
\usepackage{caption}
\usepackage{subcaption}
\usepackage{cite}
\usepackage{color}
\usepackage{setspace}
\usepackage[colorlinks=true
,urlcolor=blue
,citecolor=blue
,linkcolor=blue
,pagecolor=blue
,linktocpage=true
,pdfproducer=medialab
]{hyperref}
\usepackage{cleveref}
\usepackage[a4paper,width=17cm,height=25cm]{geometry}
\makeatletter \renewcommand{\@dotsep}{10000} \makeatother

\title{Understanding the muon anomalous magnetic moment in light of a flavor symmetry-based Minimal
	 Supersymmetric Standard Model}

\date{}

\author{Mureed Hussain\thanks{Email: mureed.hussain@sns.nust.edu.pk}~}
\author{Rizwan Khalid\thanks{Email: rizwan@sns.nust.edu.pk}}

\vspace{0.25cm}

\affil{Department of Physics, School of Natural Sciences, National University of Sciences \& Technology, H-12, Islamabad, Pakistan}

\begin{document}
\maketitle
\begin{abstract}
 We investigate whether the Minimal Supersymmetric Standard Model with the scalar masses
of the third generation distinct from the first two, in order to be able 
to accommodate the muon anomalous magnetic moment, is consistent with the latest results from
LHCb, direct collider bounds from the ATLAS and CMS experiments. In particular,
we show that this class of models allows for satisfying both the constraints from the muon $(g-2)_\mu$
experiment and various bounds from the LHC. In addition, such models can also explain the observed dark
matter relic density.
\end{abstract}

\section{Introduction}
Precision studies in B Physics provide a popular mechanism to constrain any scenario beyond 
the Standard Model (SM) in general \cite{Hurth:2006mm,Falkowski:2015zwa}, and supersymmetry 
(SUSY) in particular \cite{Arbey:2012ax,Anandakrishnan:2014nea,Misiak:1997ei}. With the Large Hadron Collider 
beauty (LHCb) experiment, we have entered a new era of precision measurement that puts 
increasingly stringent constraints on, for example, the Minimal Supersymmetric Standard Model 
(MSSM)~\cite{Arbey:2012bp,Ellis:2014cga,Buchmueller:2014yva,DeCausmaecker:2015yca}.
Constraints from rare B decays like~
 $b \rightarrow s \gamma$, $B_{s} \rightarrow \mu^{+} \mu^{-}$,
 $B_{u} \rightarrow \tau \nu_{\tau}$ are routinely used in SUSY parametric space
 studies (see for example \cite{Domingo:2007dx}). LHCb has also given quantum chromodynamics (QCD)
 form factor independent observables in connection with the semi-leptonic rare decay
 ($B_{d} \rightarrow K^{*} \mu^{+} \mu^{-}$) \cite{Aaij:2013iag}. An important result
 for this decay is the zero crossing value of the forward-backward asymmetry ($\mathcal{A}_{FB}$),
\begin{equation}\label{newB}
q^{2}_{0}\hspace{0.2cm}= \hspace{0.2cm} 4.9 \pm 0.9 \hspace{0.2cm} (GeV^{2}).
\end{equation}
In addition, a host of form factor independent observables related to
transversity amplitudes in the $B_{d} \rightarrow K^{*} \mu^{+} \mu^{-}$ decay
are now known experimentally (see Section~\ref{effectiveTheory} for details). So far, B Physics constraints do not show any $5\sigma$ deviation from the SM, and therefore, conservatively set a lower bound on new physics models.

The SM prediction for the anomalous magnetic moment of the muon, $a_{\mu}=(g-2)_{\mu}/2$ (hereafter referred 
to as the $g_\mu-2$ anomaly) 
\cite{Gohn:2016ezs}, has a $3.5 \sigma$ discrepancy with the experimental 
results \cite{BNL}. If new physics is to offer a  solution to this discrepancy, then this 
discrepancy provides both an upper and a lower bound for it. This is, of course, also true for low scale 
supersymmetry \cite{Ajaib:2015yma}. It is an interesting question to probe whether the $g_\mu-2$ 
anomaly can be resolved consistently with the indirect probes via B-physics in regions allowed
 by the direct searches for SUSY particles.

The discovery of the Higgs boson with a mass $\sim125$ GeV~\cite{:2012gk,:2012gu} has 
important consequences, in general, for low scale supersymmetry. In the MSSM such a heavy Higgs mass 
requires either a large, ${\cal O} (\mathrm{few}-10)$ TeV  stop squark mass, or alternatively, a relatively large 
soft supersymmetry  breaking (SSB) trilinear $A_t$-term, along with a stop squark mass of around a 
TeV  \cite{Heinemeyer:2011aa}. The constraints coming from the Higgs boson mass are  particularly 
stringent regarding the  sparticle spectrum if we assume universal boundary conditions on the soft
SUSY breaking (SSB) parameters at some high 
energy scale (typically $M_{GUT} \sim10^{16}GeV $). Such universal boundary conditions 
for the SSB terms are well motivated in minimal scenarios of gravity\cite{Baer:2011ab} or gauge mediation~\cite{Draper:2011aa}. 
For instance, in the simple version of gravity or gauge mediation scenarios with universal 
scalar and gaugino mass terms, it is difficult to simultaneously explain the 
observed Higgs boson mass and resolve the $g_\mu-2$ anomaly. There have been several recent 
attempts to reconcile this presumed tension between the muon $g_\mu-2$ and the 
125 GeV Higgs mass~\cite{Zhu:2016ncq}. 

In this paper, we use the framework of flavor symmetry based Minimal Supersymmetric 
Standard Model (sMSSM) suggested recently \cite{Babu:2014sga,Babu:2014lwa}, consisting of seven 
phenomenological parameters which describe SUSY breaking. We show that it is possible to explain the 
$g_\mu-$~$2$ 
anomaly and the Higgs boson mass simultaneously, along with the observed dark matter 
abundance in addition to satisfying direct mass bounds on sparticles and constraints from LHCb.
In doing so, we follow
the approach of~\cite{Babu:2014lwa}. However, we have incorporated the constraints from the
$B_{d}\to K^{*} \mu^{+} \mu^{-}$ channel which were not studied  in~\cite{Babu:2014lwa} and in addition,
we have used updated values of all constraints.

In Section~\ref{effectiveTheory}, we give the effective theory concerning semi-leptonic
weak decays and discuss several experimental observables that we incorporate in our study.
In Section~\ref{scanProc}, we describe details of the sMSSM parameters along with our 
scanning procedure and constraints.
In Section~\ref{results}, we discuss our results after which we conclude in Section~\ref{conclusion}.


\section{Effective theory of semi-leptonic B decays}
\label{effectiveTheory}
Despite being successful probes of new physics, the decays $b \rightarrow s \gamma$,
$B_s \rightarrow \mu^{+} \mu^{-}$ and $B_{u} \rightarrow \tau \nu_{\tau}$ have a small number of 
observables like CP asymmetries and branching ratios.
The constraints on MSSM from the branching ratio of $b \rightarrow s \gamma$
and its time dependent CP asymmetries  are studied in 
\cite{Ellis:2007fu,Altmannshofer:2009ne,Bartl:2001ka}. The rare decay $B_s \rightarrow \mu^{+} \mu^{-}$
which is helicity suppressed in the SM and so contributes by additional diagrams in the MSSM is
also used for constraining the MSSM parameter space as in
\cite{Arbey:2012ax,Mahmoudi:2012uk,Mahmoudi:2012un}. The rare decay
$B_{d} \rightarrow K^{*} \mu^{+} \mu^{-}$ provides along with the differential branching ratio as a
function of square of the dilepton invariant mass ($q^{2}$), observables such as $K^{*}$ longitudinal polarization
fraction ($F_{L}$) and forward-backward asymmetry ($\mathcal{A}_{FB}$), where
\begin{align}
\label{AFB}
\begin{split} 
F_{L}&= \frac{\Gamma_{L}}{\Gamma}~,
\\
\mathcal{A}_{FB} &=\frac{N_F - N_B}{N_F + N_B}.
\end{split}
\end{align}
$\Gamma_L$ is the decay rate of $B_{d} \rightarrow K^{*} \mu^{+} \mu^{-}$ when $K^{*}$ is 
longitudinally polarized and 
$\Gamma$ is the total decay rate.
Likewise, $N_F$ is the number of events in which $\mu^-$($\mu^+$) is moving in the \emph{forward} direction 
with respect to $B_{d}$($\bar B_{d}$) in the dilepton 
rest frame, and $N_B$ is the corresponding number of events in the \emph{backward} direction. In 
terms of the differential cross-section $\frac{d\sigma}{d\Omega}$, $\mathcal{A}_{FB}$ is given as
\begin{equation}
\mathcal{A}_{FB} = \frac{\int\limits_{0}^{\pi} d\Omega \frac{d\sigma}{d\Omega}-\int\limits^{0}_{-\pi}d\Omega \frac{d\sigma}{d\Omega}}{\int \limits^{\pi}_{-\pi} d\Omega \frac{d\sigma}{d\Omega}}.
\end{equation}
In addition, the zero crossing of this $\mathcal{A}_{FB}$ given in Eq.~\eqref{newB} is also important as it puts 
constraints on a variety of models~\cite{Mahmoudi:2014iia}. 

Semi-leptonic part of the effective Hamiltonian which is most sensitive to the $b \rightarrow s \ell
\ell$ \cite{Buras:1998raa,Mahmoudi:2014mja} decay is composed of radiative and dileptonic operators
and is given by
\begin{equation}\label{EffectiveH}
\mathcal{H}^{{\text{\tiny sl}}}_{{\text{\tiny eff}}}=-\frac{4G_{F}}{\sqrt{2}}V_{{\text{\tiny tb}}}
V_{{\text{\tiny ts}}}^{*} \bigg [\sum_{i=7,9,10} (\mathcal{C}_{i} \mathcal{O}_{i} +\mathcal{C}^{'}_{i}
\mathcal{O}^{'}_{i} ) + \sum_{i=1,2}(\mathcal{C}_{{\text{\tiny $\mathcal Q$}}_{i}} \mathcal{Q}_{i} +
\mathcal{C}^{'}_{{\text{\tiny $\mathcal Q$}}_{i}} \mathcal{Q}^{'}_{i}) \bigg ],
\end{equation}
where the operators 
in $\mathcal{H}^{{\text{\tiny sl}}}_{{\text{\tiny eff}}}$ are,
\begin{equation}
\begin{array}{rcl}
\vspace{0.2cm}
 &\mathcal{O}^{( \prime ) }_{7}& = (\frac{e}{16\pi^2}) m_{b}[\overline{s}\sigma^{\mu \nu}P_{R(L)}b]F_{\mu \nu},
\hspace{0.2cm}
\mathcal{O}^{(\prime)}_{9} = (\frac{e^{2}}{16\pi^2}) [\overline{s}\gamma^{\mu}P_{L(R)}b][\overline{\ell}
\gamma_{\mu}\ell], \\
\vspace{0.2cm}
 &\mathcal{O}^{(\prime)}_{10}& = (\frac{e^{2}}{16\pi^2}) m_{b}[\overline{s}\gamma^{\mu}P_{L(R)}b][\overline{\ell}
\gamma_{\mu}\gamma_{{\text{\tiny 5}}}\ell], \\
\vspace{0.2cm}
 &\mathcal{Q}^{(\prime)}_{1}& = (\frac{e^{2}}{16\pi^2}) m_{b}[\overline{s}P_{R(L)}b][\overline{\ell} \ell],
\hspace{0.65cm}
\mathcal{Q}^{(\prime)}_{2} = (\frac{e^{2}}{16\pi^2}) m_{b}[\overline{s}P_{R(L)}b][\overline{\ell}
\gamma_{{\text{\tiny 5}}} \ell].
\end{array}
\end{equation}
The four-body final state ($B \to K^{*}  \ell^{+}\ell^{-}$ (where $K^{*} \to K \pi$)) differential decay
distribution provides a variety of experimental constraints. This differential decay distribution depends on the 
following kinematic variables:
\begin{itemize}
\item[$q^{2}$:] The invariant mass square of lepton.
\item[$\theta_{\ell}$:] Angle between the directions of flight of the $\ell^{+}(\ell^{-})$ and the
$B$ meson in the dilepton rest frame.
\item[$\theta_{K}$:] Angle between directions of flight of kaon (K) and the $B$ meson in
the rest frame of $K^{*}$.
\item[$\phi$:] The azimuthal angle between the planes of lepton pair and the $K \pi $ system.
\end{itemize}
In terms of these kinematic variables, the differential decay rate is
\begin{align}
\frac{d^4 \Gamma({B}_{d})}{dq^{2}d\cos\theta_{K}d\cos\theta_{\ell}d\phi}=&
\frac{9}{32\pi}[J_{1s}\sin^{2}\theta_{K} + J_{1c}\cos^{2}\theta_{K}+(J_{2s}\sin^{2}\theta_{K}
+J_{2c}\cos^2\theta_{K})\cos2\theta_{\ell} \nonumber \\
&+ J_{3}\sin^{2}\theta_{K}\sin^{2} \theta_{\ell}\cos2\phi+J_{4}\sin2\theta_{K}\sin2\theta_{\ell}\cos\phi \nonumber\\
&+J_{5}\sin2\theta_{K}\sin\theta_{\ell}\cos\phi
+(J_{6s}\sin^{2}\theta_{K} + J_{6c}\cos^{2}\theta_{K})\cos\theta_{\ell}\nonumber \\
&+J_{7}\sin2\theta_{K}\sin\theta_{\ell}sin\phi + J_{8}\sin2\theta_{K}\sin2\theta_{\ell}\sin\phi \nonumber \\
&+ J_{9}\sin^{2}\theta_{K}\sin^{2}\theta_{\ell}\sin2\phi].
\end{align}
The coefficients $J_{i}$ depend on the transversity amplitudes (decay amplitudes in which the
particles' spins are projected normal to the reaction plane) and their explicit form is
given in \cite{Matias:2012xw}.
The fully accessible phase space is bounded from the kinematics by
\begin{equation*}
 4m_{\ell ^ {2}} \le q^{2} \le (M_{B} - m_{K^{*}})^{2} ,\hspace{0.5cm} -1 \le \cos\theta_{\ell} \le 1,
\hspace{0.5cm}-1 \le \cos\theta_{K} \le 1, \hspace{0.5cm} 0 \le \phi \le 2 \pi.
\end{equation*}

These $J_{i}(q^2)$ integrated in different $q^2$ bins form the basic observables for this decay. To minimize
the hadronic uncertainties some optimized (form factor independent) observables can be constructed by taking appropriate
ratios of these $J_{i}'s$. 
All of these observables from
$B_{d} \rightarrow K^{*} \mu^{+} \mu^{-}$ can be measured at the LHCb as a function of $q^{2}$
and this decay proves to be very important one for constraining the new physics scenarios~\cite{Hurth:2016fbr}.
The optimized observables denoted by $P_i$ are,

\begin{eqnarray}\label{Pi}
&&\langle P_{1}\rangle_{{\text {\tiny bin}}}=\frac{1}{2}\frac{\int_{{\text{\tiny bin}}}dq^{2}[J_{3}+\overline{J}_{3}]}
{\int_{{\text{\tiny bin}}}dq^{2}[J_{2s}+\overline{J}_{2s}]},
\hspace{0.8cm}
\langle P_{2}\rangle_{{\text {\tiny bin}}}=\frac{1}{8}\frac{\int_{{\text{\tiny bin}}}dq^{2}[J_{6s}+\overline{J}_{6s}]}
{\int_{{\text{\tiny bin}}}dq^{2}[J_{2s}+\overline{J}_{2s}]}, \nonumber \\
&&\langle P^{'}_{4}\rangle_{{\text {\tiny bin}}}=\frac{1}{\mathcal{N}^{'}_{{\text{\tiny bin}}}}
\int_{{\text{\tiny bin}}}dq^{2}[J_{4}+\overline{J}_{4}],
\hspace{0.6cm}
\langle P^{'}_{5}\rangle_{{\text {\tiny bin}}}=\frac{1}{2\mathcal{N}^{'}_{{\text{\tiny bin}}}}
\int_{{\text{\tiny bin}}}dq^{2}[J_{5}+\overline{J}_{5}], \nonumber \\
&&\langle P^{'}_{6}\rangle_{{\text {\tiny bin}}}=\frac{-1}{2\mathcal{N}^{'}_{{\text{\tiny bin}}}}
\int_{{\text{\tiny bin}}}dq^{2}[J_{7}+\overline{J}_{7}],
\hspace{0.4cm}
\langle P^{'}_{8}\rangle_{{\text {\tiny bin}}}=\frac{1}{\mathcal{N}^{'}_{{\text{\tiny bin}}}}
\int_{{\text{\tiny bin}}}dq^{2}[J_{8}+\overline{J}_{8}], \nonumber
\end{eqnarray}
where $\overline{J}_{i}$ correspond to the decay $\overline{B} \to \overline{K}^{*} \mu^{-}\mu^{+}$ and
 the normalization factor is
\begin{equation}
\mathcal{N}^{\hspace{0.04cm}'}_{{\text{\tiny bin}}}=\sqrt{-\int_{{\text{\tiny bin}}}dq^{2}[J_{2s}+\overline{J}_{2s}]
\int_{{\text{\tiny bin}}}dq^{2}[J_{2c}+\overline{J}_{2c}]}. \nonumber
\end{equation}
A comprehensive study of this preferable choice of observables in light of results from LHCb has been done
in \cite{Descotes-Genon:2013zva}.

\section{Flavor symmetry-based MSSM parameter space, scanning procedure and constraints}
\label{scanProc}
We use the sMSSM described in \cite{Babu:2014sga} as the basis of our study.
In the sMSSM, the SSB Lagrangian is consistent with two symmetries (a)
a GUT symmetry such as $SO(10)$ and (b) a non-abelian flavor symmetry of gauge
origin that acts on the three families with either a {\bf 2+1} or a {\bf 3} family assignment.
The GUT scale symmetry reduces the MSSM parameters, for example, for gauginos the $SO(10)$ symmetry reduces the number of parameters from three to one. It also suggests that all members of a family would have a common soft mass, 
as they are unified into a {\bf 16-}plet of $SO(10)$.

The non-abelian flavor symmetry suppresses the SUSY mediated flavor
changing neutral current~(FCNC) processes mediated by SUSY particles.
This FCNC suppression is required here as we
want our model to be compatible with GUT and also have different masses for the sfermions families.
Any symmetry like $SU(2)_f$ which has a doublet representation can be used as flavor symmetry but such
symmetry will contain new sources of flavor violation, arising from the $D$-terms
which split the masses of superparticles within a given multiplet after SUSY
breaking\cite{Kawamura:1994ys}. An interchange symmetry has been suggested~\cite{Babu:2014sga} that
would set these $D$-terms to zero.
So, for the flavor symmetry $SU(2)_f$, soft masses of the scalars for the $1^{\rm st}$ and
$2^{\rm nd}$ family are in
a doublet under $SU(2)_f$ (${\mathbf 16}_{1}, {\mathbf 16}_{2}$) while the third family
is a singlet (${\mathbf 16}_{3}$) under $SU(2)_f$.

Together these two symmetries reduce the 15 soft squared mass parameters of the 15
chiral sfermions of the
MSSM to just two. We consider the SUSY phenomenology of sMSSM to be described 
by seven parameters
\begin{equation}
m_{1,2},~ m_3,~ M_{1/2},~ A_{0},~ tan\beta,~ m_{H_{u}},~m_{H_{d}}
\end{equation}
where $m_{1,2}$ is the common mass parameter of the first two family sfermions, $m_{3}$ is the mass parameter of the third family
sfermions, $M_{1/2}$ is the unified gaugino mass parameter, $A_{0}$ is the unified
trilinear coupling parameter and $\tan\beta$ is the ratio of the vacuum expectation values of the two Higgs doublets. Finally, $m_{H_{u}}$ and $m_{H_{d}}$ are the corresponding SSB Higgs mass parameters for the two Higgs doublets which are set separately from any sfermions mass paramenters.
We employ the SOFTSUSY-3.5.2 package \cite{SOFTSUSY},
which calculates the sparticle spectrum in the CP-conserving MSSM with a full
flavor mixing structure to perform the random scans over the parametric space. This program
solves the renormalization group equations with boundary conditions on the SSB terms specified at $M_{GUT}$.
Weak scale gauge couplings and fermion mass data are used as a boundary condition at $M_Z$ (the Z boson mass).
The sMSSM parametric space that we have scanned is,

\begin{equation}\label{parsNUHMII}
\begin{array}{rcl}
0 ~\le &m_{1,2}& \le ~3 ~\text{TeV}, \\
0 ~\le &m_{3}& \le ~3 ~\text{TeV}, \\
0 ~\le &M_{1/2}& \le ~ 3 ~\text{TeV},\\
-3~\le &A_{0}/m_{3}& \le ~ 3,  \\
0~\le &\tan\beta & \le ~60,\\
0~\le &m_{H_{u}}& \le ~5 ~\text{TeV},\\
0~\le &m_{H_{d}}& \le ~5 ~\text{TeV},\\
&\mu&>~ 0.
\end{array}
\end{equation}

In other words, we basically consider the parametric space of the non-universal Higgs model of type II (NUHM-II)~\cite{Ellis:2002iu}, having independent parameters for up-type and down-type Higgs $m_{H_{u}}$ and $m_{H_{d}}$,
but with split masses of sfermions. Later in this article we will also discuss the scan over the
parametric space of NUHM-I in which these two Higgs mass parameters are set equal to each other.

After the generation of SUSY LesHouches Accord (SLHA)~\cite{Skands:2003cj} file via SOFTSUSY and hence the sparticle spectrum, we use the SUPERISO package~\cite{Mahmoudi:2008tp} to calculate different B Physics observables.

Branching ratios of the $B$ decays $B_{s} \rightarrow \mu^{+} \mu^{-}$, $b \rightarrow s \gamma$ and $B_{u} \rightarrow \tau \nu_{\tau} $ have been used to constrain the
parametric space of different MSSM models~\cite{Ellis:2007fu,Altmannshofer:2009ne,Bartl:2001ka,Mahmoudi:2012uk,Mahmoudi:2012un}.
LHC also provides lower bounds on sparticle masses~\cite{Aad:2012fqa, Cervelli:2015pga, Aad:2015eda,Abreu:2000kia, Abdallah:2003xe, Dreiner:2009ic}. The Muon g-2 
Collaboration \cite{Bennett:2006fi} provides a significantly precise value of the anomalous magnetic
moment of the muon. The difference between the experimental value of $g_{\mu} -2$ and its theoretical value calculated in the Standard Model is defined as $
 \Delta a_{\mu} = a_{\mu}^{\text{exp}} - a_{\mu}^{\text{SM}}.$
 
We apply these constraints along with constraints on the SM-like Higgs boson mass, on the
parametric space of sMSSM. These constraints are
\begin{equation}\label{gluino}
\left.
\begin{array}{rcl}
m_{\tilde{g}} &>& 1900.0 \hspace{0.2cm} \text{GeV} , \\
m_{\tilde{q}}  &>&  1600.0 \hspace{0.2cm}  \text{GeV},\\
m_{\tilde{\chi}^{0}_{1}} & >& 46 \hspace{0.2cm}  \text{GeV},\\
m_{\tilde{\chi}^{0}_{2}} & >& 670.0 \hspace{0.2cm}  \text{GeV} 
\hspace{0.2cm}(m_{\tilde{\chi}^{0}_{1}} < 200 \hspace{0.2cm}  \text{GeV}),\\
m_{\tilde{\chi}^{0}_{2}} & >& 116.0 \hspace{0.2cm}  \text{GeV},\\
m_{\tilde{\chi}^{\pm}_{1}} & >& 103.5 \hspace{0.2cm}  \text{GeV} 
\hspace{0.2cm}(m_{\tilde{\nu}} > 300 \hspace{0.2cm}\text{GeV}),\\
m_{\tilde{\chi}^{\pm}_{1}} & >& 94.0 \hspace{0.2cm}  \text{GeV}
\end{array}
\right\}
\end{equation}

\begin{equation}\label{higgs}
123.0 \hspace{0.1cm} \text{GeV} \hspace{0.2cm} \le m_{h^{0}} \le \hspace{0.2cm} 127.0 \hspace{0.1cm} \text{GeV}.
\end{equation}
 Constraints from branching ratios of rare decays
$b \to s \gamma$, $B_s \to \mu^+ \mu^-$ and $B_{u} \to \tau \nu_{\tau}$,
\begin{equation}\label{oldB}
\left.
\begin{array}{rcl}
BR(B_s \rightarrow \mu^{+}\mu^{-})& = & (2.9 \pm 0.7)\times 10^{-9}, \\
BR(b \rightarrow s \gamma) & = & (3.43 \pm 0.22 )\times 10^{-4},  \\
\frac{BR(B_{u} \rightarrow \tau \nu_{\tau})_{\text{MSSM}}}{BR(B_{u} \rightarrow \tau \nu_{\tau})_{\text{SM}}} &=& 1.13 \pm 0.43 .
\end{array}
\right\}
\end{equation}
and constraints from optimized observable $P_{i}'s$ and $q_{0}^{2}$ of $\mathcal{A}_{FB}$ for rare decay $B_{d} \to K^{*} \mu^+ \mu^-$, given in \cite{Mahmoudi:2014mja}, are applied.  

We calculate these observables for each point and then compare them to experimental results by calculating $\chi^2$ given as:
\begin{eqnarray}
\chi^2 = &&\sum\limits_{\text{bins}}\bigg[\sum\limits_{{i,j}\in (B \to K^{*} \mu ^+ \mu^- obs.)} 
			(O^{\text{exp}}_{i}- O^{\text{th}}_{i})
         (\sigma^{(\text{bin})})^{-1}(O^{\text{exp}}_{j}- O^{\text{th}}_{j})\bigg]\\ 
         &&+ \sum\limits_{k \in (\text{other B Physics obs.})}\frac{(O^{\text{exp}}_{k}- O^{\text{th}}_{k})}{(\sigma^{\text{exp}}_{k}- \sigma^{\text{th}}_{k})}.
\end{eqnarray}\label{chiSquare}
Each data point has a particular $p\text{-value}$ which is then used to measure the confidence 
level (CL) by $(1-p) \times 100$. We have
implemented the $\chi^2$ analysis technique used by \cite{Hurth:2016fbr} and refer the same for more details. 
Finally, we apply constraints from $g_{\mu}-2$ anomaly $1$-$\sigma$ range
\begin{equation}\label{g-2}
\Delta a_{\mu} = 28.6 \pm 8.0 \times 10^{-10}.
\end{equation}
\section{Results}
\label{results}
\subsection{sMSSM with NUHM-II}
The Feynmann diagrams for the $g_{\mu} -2 $ calculations involve neutralino-smuon exhange or chargino-sneutrino exchange. Diagrams for rare B decays contain the third family sparticles in the loops along with charged Higgs; so we have made plots for these sparticles masses. Also there are squarks and gluino mass bounds from experiments like ATLAS and CMS. We have made those plots so that we can understand the mass bounds imposed by flavor physics data of B decays and the $g_{\mu}-2$ anomaly.

\begin{figure}[t]
	\captionsetup{width=0.90\textwidth}
	\begin{subfigure}{.5\textwidth}
		\includegraphics[width=8cm,height=6cm]{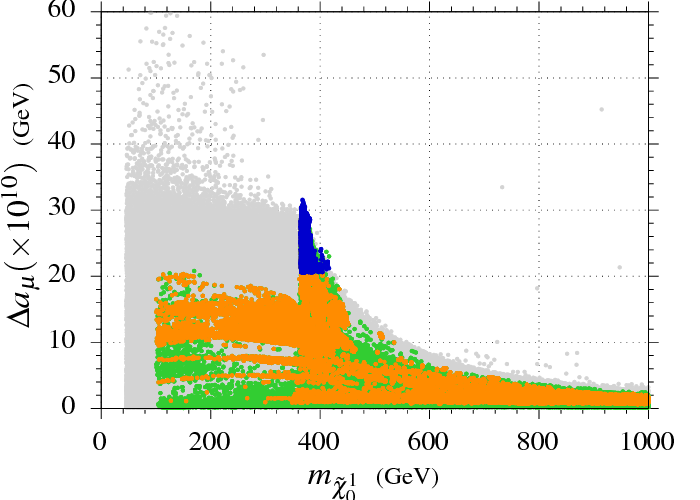}
		\label{fig:sfig11}
	\end{subfigure}
	\begin{subfigure}{.5\textwidth}
		\includegraphics[width=8cm,height=6cm]{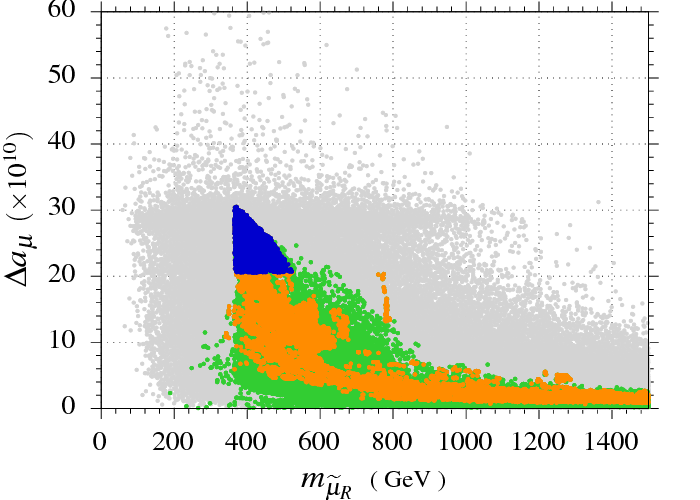}
		\label{sfig12}
	\end{subfigure}
	\begin{subfigure}{.5\textwidth}
		\includegraphics[width=8cm,height=6cm]{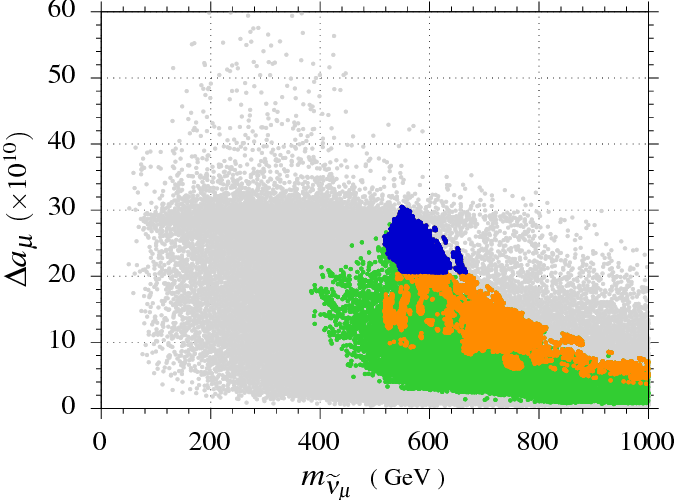}
		\label{fig:sfig13}
	\end{subfigure}
	\begin{subfigure}{.5\textwidth}
		\includegraphics[width=8cm,height=6cm]{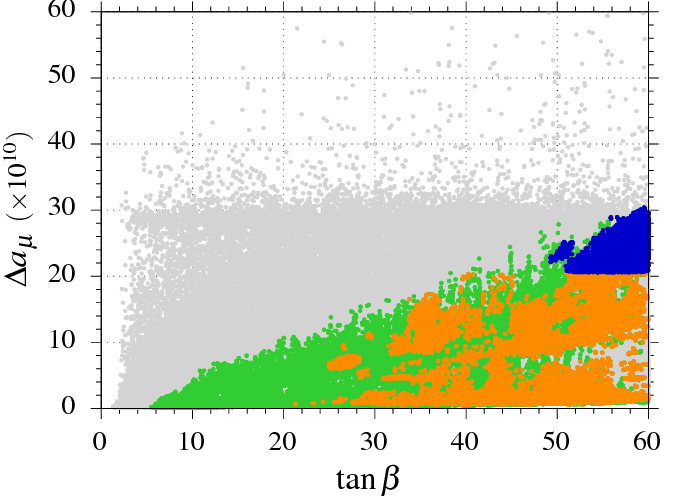}
		\label{fig:sfig14}
	\end{subfigure}
	\caption{All points satisfy the REWSB and neutralino LSP conditions.
		Green points form a subset of grey points and satisfy experimental lower-mass 
		bounds on neutralinos, charginos, gluinos and squarks. These points also satisfy theoretical mass bounds on  
	    SM-like Higgs particle. 
		Orange points are a subset of green points and lie within 95$ \% $ confidence level
		region of all B Physics observables from branching ratios of $b \to s \gamma$, 
		$B_u \to \tau \nu_{\tau}$, $B_s \to \mu^+ \mu^- $ in addition to 
	    zero crossing of $\mathcal{A}_{FB}$ and seven other constraints, 
		on optimized observables {$P_{i}$'s}, from $B_{d} \to K ^{*} \mu^{+} \mu^{-}$
		 as given in \cite{Mahmoudi:2014mja}.
		Blue points form subset of orange points and lie within the $1$-$\sigma$ allowed range of the
		$g_{\mu}-2$ anomaly.}
	\label{fig:fig1}
\end{figure}

We first present our results of the scan over the parametric space given in
Eq.~\eqref{parsNUHMII}. In Fig.~\ref{fig:fig1}, we present results in the
$\Delta a_{\mu}$ - $m_{\tilde{\chi}^{0}_{1}}$, $\Delta a_{\mu}$ - $m_{\tilde{\mu}_{R}}$,
$\Delta a_{\mu}$ - $m_{\tilde{\nu}_{\mu}} $ and $\Delta a_{\mu}$ - $\tan \beta$ planes. 
$m_{\tilde{\chi}^{0}_{1}}$ is the mass of the lightest neutralino, $m_{\tilde{\mu}_{R}}$ is the mass of the SUSY partner of right handed muon and $m_{\tilde{\nu}_{\mu}} $ is the mass of SUSY partner of muon-neutrino. Grey points are consistent with radiative electroweak symmetry breaking (REWSB) and these points also satisfy the condition of the neutralino being the lightest supersymmetric particle (LSP).

Green points form a subset of grey
points and satisfy the neutralinos, charginos, gluinos and squarks mass constraints given in Eq.~\eqref{gluino}. These points also satisfy the Higgs mass bounds given in Eq.~\eqref{higgs}
Orange points form a subset of green
points and they have 95$\%$ CL for all B Physics constraints discussed in last section.
Blue points are the subset of orange points that satisfy the $g_{\mu}-2$ constraints given in Eq.~\eqref{g-2}.

We can, from the $\Delta a_{\mu}$ - $m_{\tilde{\chi}^{0}_{1}}$ plane in Fig.~\ref{fig:fig1}, infer that sparticles and Higgs mass bounds put a lower limit of $\sim100$ GeV on $m_{\tilde{\chi}^{0}_{1}}$ which is also consistent with B Physics constraints. We can also see that in order to satisfy the $g_{\mu}-2$ anomaly constraint, the neutralino should be heavier than 360 GeV and lighter than 420 GeV. This range is also consistent with the constraints from LHCb data on rare B decays in addition to all the other constraints mentioned in previous section.  

It can be seen in $\Delta a_{\mu}$ - $m_{\tilde{\mu}_{R}}$ plane of Fig.~\ref{fig:fig1} that the sparticles mass bounds put a lower bound of 200 GeV on $m_{\tilde{\mu}_{R}}$ while B Physics constraints push this lower bound to $\sim340$ GeV. The $g_{\mu}-2$ anomaly correction due to sMSSM inversely depends on square of smuon mass so satisfying this constraint applies an upper bound on smuon mass and we find the mass of $\tilde{\mu}_{R}$ to be in the range $360$ GeV $\lesssim m_{\tilde{\mu}_{R}} \lesssim 550$ GeV from the $g_{\mu}-2$ constraint.

From $\Delta a_{\mu}$ - $m_{\tilde{\nu}_{\mu}} $ plane we can see that the sparticles and Higgs mass bounds put a lower limit of 380 GeV on smuon-neutrino mass. B Physics constraints enhance this lower bound to 500 GeV which is also consistent with $g_{\mu}-2$ anomaly constraint.  The $g_{\mu}-2$ anamoly  constraint also puts an upper bound of 700 GeV on smuon-neutrino mass.

In the $\Delta a_{\mu}$ - $\tan\beta$ plane of Fig.~\ref{fig:fig1}. We see that $\tan\beta < 5$ is not allowed (mainly due to Higgs mass bounds). The hard cut is from squarks and gluino mass bounds. B Physics observables do not allow $\tan \beta <20$ for this particular model. $\tan \beta$ behaves linearly in the function describing the $g_{\mu}-2$ anomaly correction in sMSSM so $g_{\mu} -2$ constraint pushes this limit to $\tan \beta \sim 48$. This intense change in lower bounds explain dependence of all constraints on $\tan \beta$.
\begin{figure}[t]
	\captionsetup{width=0.90\textwidth}
	\begin{subfigure}{.5\textwidth}
		\centering
		\includegraphics[width=8cm,height=6cm]{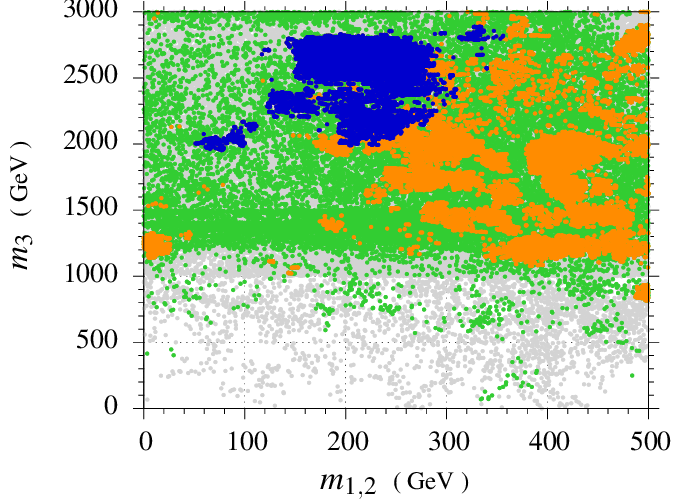}
		\label{fig:sfig21}
	\end{subfigure}
	\begin{subfigure}{.5\textwidth}
		\centering
		\includegraphics[width=8cm,height=6cm]{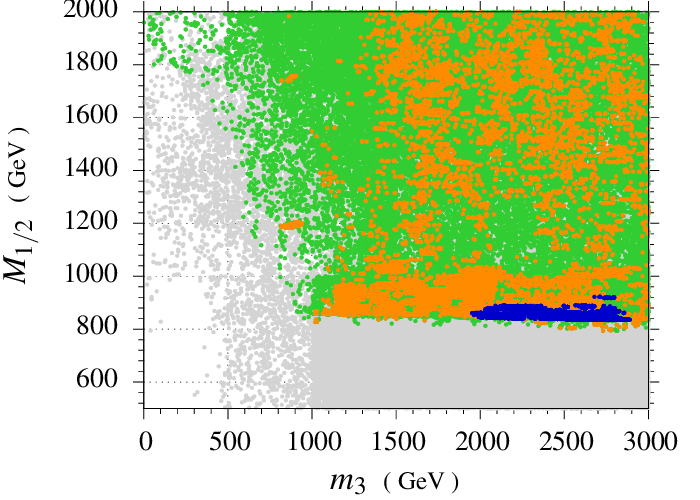}
		\label{fig:sfig22}
	\end{subfigure}\\
	\caption{Plots in $m_{3}$ - $m_{1,2}$ and $M_{1/2}$ - $m_{1,2}$ planes. Color coding is the same as in Fig.~\ref{fig:fig1}. Clearly CMSSM($m_{1,2} = m_{3}$) is ruled out by $g_{\mu}-2$ constraint.} 	
	\label{fig:fig2}
\end{figure}

In Fig.~\ref{fig:fig2}, $m_{3}$ - $m_{1,2}$  plane, we can see that in order to satisfy the $g_{\mu}-2$
constraint along with other bounds, $m_{3}/m_{1,2}\gtrsim 9$. This is because a heavy $m_3$ is needed 
to satisfy, for example, the Higgs mass bound and direct collider bounds on the stop. Likewise, 
a relatively small $m_{1,2}$ ensures that the muon and tauon sneutrino are not too heavy and can 
contribute to the muon $g_\mu-2$. We can see that in order to satisfy the $g_\mu-2$ constraint, the upper bound on $m_{1,2}$ is 360 GeV and for that the lower bound on $m_{3}$ is $\sim1900$~GeV.

Since the $g_{\mu}-2$ constraint requires gauginos (bino or wino) to be lighter, we present our 
results in the $M_{1/2}$ - $m_{1,2}$ plane of Fig.~\ref{fig:fig2}. Gaugino masses are unified in our model (given by one parameter ``$M_{1/2}$") and they are restricted to 900 GeV from the $g_{\mu}-2$ anomaly constraint. The sharp cut at $M_{1/2}$ $\sim800$ GeV is essentially due to the heavy bound on gluino mass.

\begin{figure}[H]
	\begin{subfigure}{.5\textwidth}
				\centering
		\includegraphics[width=8cm,height=6cm]{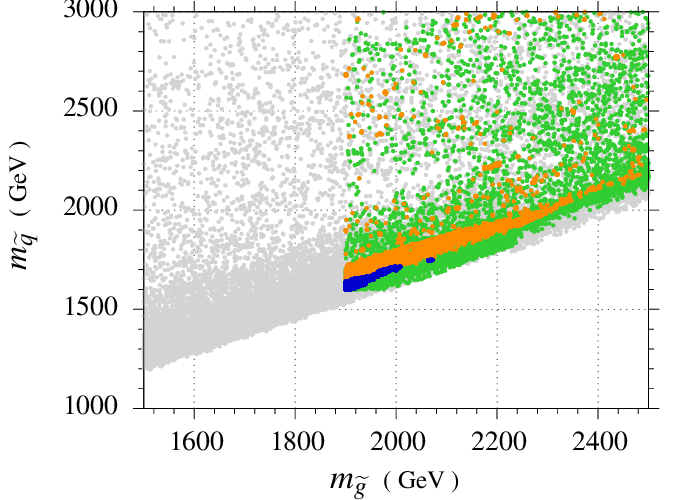}
		\label{fig:sfig31}
	\end{subfigure}
	\begin{subfigure}{.5\textwidth}
		\centering
		\includegraphics[width=8cm,height=6cm]{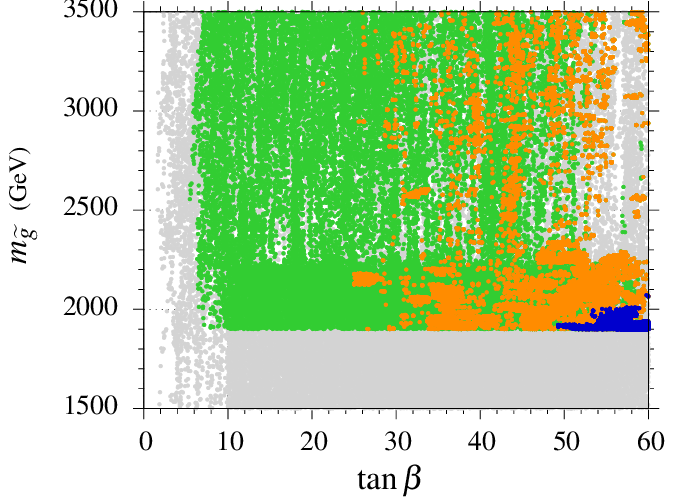}
		\label{fig:sfig32}
	\end{subfigure}
	\caption{Plots in the $m_{\tilde{q}}$ - $m_{\tilde{g}}$ and
		$m_{\tilde{g}}$ - $\tan\beta$ planes, color coding is the same as in Fig.
		\ref{fig:fig1}.}
	\label{fig:fig3}
\end{figure}

As we have mentioned in the previous section that there are mass bounds on sparticles from the CMS and ATLAS experiments so it will be interesting to check the impact of B Physics and $g_{\mu} - 2$ constraints on their remaining mass ranges. We plot the $m_{\tilde{q}}$ (the lightest of the first two generation squarks) verses the gluino mass $m_{\tilde{g}}$ in the left panel of Fig.~\ref{fig:fig3}. We show that there is an upper bound of $\sim1800$ GeV on the light squark masses which arises from the fact that the contribution from the smuon is important for resolving the muon $g_\mu-2$ anomaly in the MSSM. Likewise, the lower bound is imposed by the direct mass bounds on $m_{\tilde{q}}$ coming from LHC.

\begin{figure}[H]
	\captionsetup{width=0.90\textwidth}
	\begin{subfigure}{.5\textwidth}
		\centering
		\includegraphics[width=8cm,height=6cm]{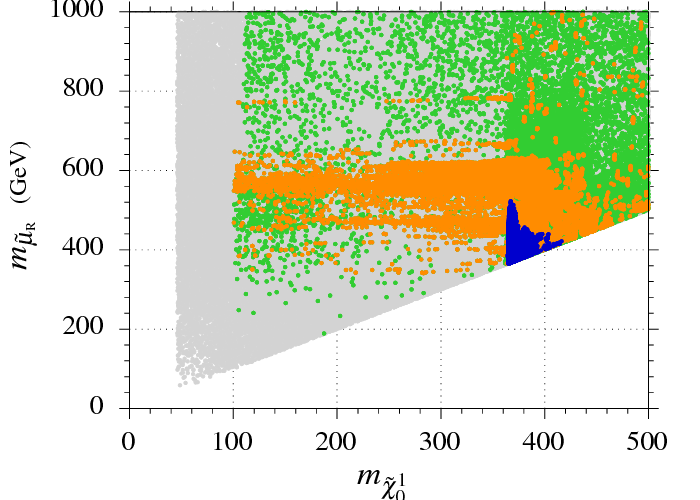}
		\label{fig:sfig41}
	\end{subfigure}
	\begin{subfigure}{.5\textwidth}
		\includegraphics[width=8cm,height=6cm]{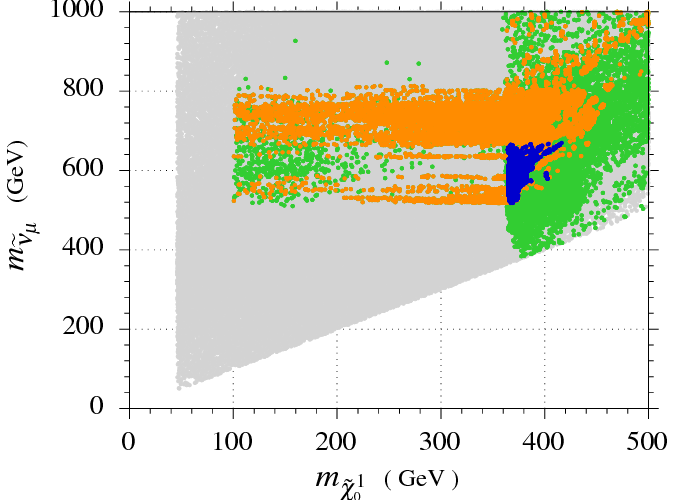}
		\label{fig:sfig42}
	\end{subfigure}
	\begin{subfigure}{.5\textwidth}
		\includegraphics[width=8cm,height=6cm]{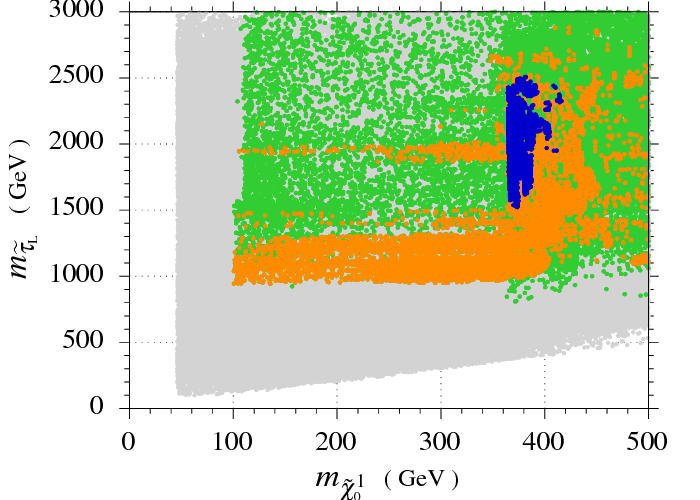}
		\label{fig:sfig43}
	\end{subfigure}
	\begin{subfigure}{.5\textwidth}
		\includegraphics[width=8cm,height=6cm]{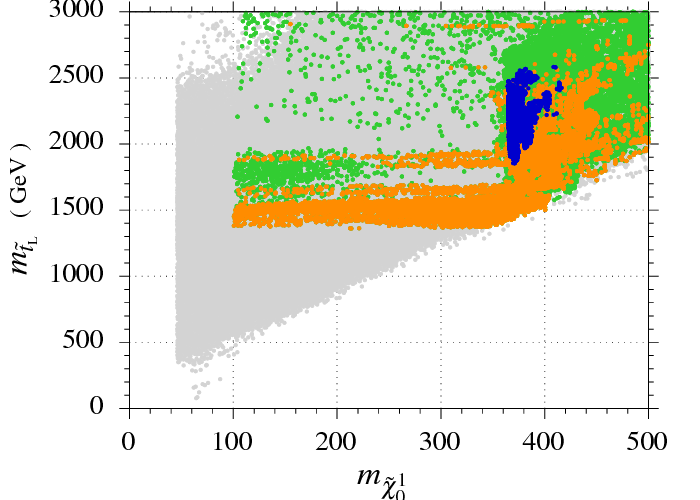}
		\label{fig:sfig44}
	\end{subfigure}
	\caption{Plots in the $m_{\tilde{\mu}_{R}}-m_{\tilde{\chi}^{0}_{1}}$,
		$m_{\tilde{\nu}_{\mu}}-m_{\tilde{\chi}^{0}_{1}}$,
		$m_{\tilde{\tau}_{\text{\tiny L}}}-m_{\tilde{\chi}^{0}_{1}}$,
		$m_{\tilde{t}_{\text{\tiny L}}}-m_{\tilde{\chi}^{0}_{1}}$ planes, color coding is the same as in Fig.~\ref{fig:fig1}.}
	\label{fig:fig4}
\end{figure}

Plot in the $m_{\tilde{g}}$ - $\tan\beta$ plane of Fig.~\ref{fig:fig3} shows that 
the upper limit on $m_{\tilde{g}}$ for our chosen model is $\sim2100$ GeV.
Next, in Fig.~\ref{fig:fig4}, we show our results in the $m_{\tilde{\mu}_{R}}$-$m_{\tilde{\chi}^{0}_{1}}$,
$m_{\tilde{\nu}_{\mu}}$-$m_{\tilde{\chi}^{0}_{1}}$,
$m_{\tilde{\tau}_{\text{\tiny L}}}$-$m_{\tilde{\chi}^{0}_{1}}$ and
$m_{\tilde{t}_{\text{\tiny L}}}$-$m_{\tilde{\chi}^{0}_{1}}$ planes, where $m_{\tilde{\nu}_{\mu}}$ is the mass of 
the SUSY partner of muon neutrino, $m_{\tilde{\tau}_{\text{\tiny L}}}$ is the mass of the SUSY partner of
left-handed tauon and $m_{\tilde{t}_{\text{\tiny L}}}$ is the mass of the SUSY partner of left-handed top quark.

We see that the co-annihilation scenario involving the smuon is compatible with B Physics constraints and the $1$-$\sigma$ range of $g_{\mu}-2$, so we expect the possibility of neutralino dark matter\footnote{We provide an example of the ${\tilde{\mu}_{R}}$ co-annihilation with the corresponding relic densities in Table~\ref{t1}.}. This co-annihilation occurs in the neutralino mass range from 350 GeV to nearly 420 GeV.

We can see from Fig.~\ref{fig:fig4} that the $g_{\mu}-2$ anomaly constraint 
play a crucial role in constraining the parameter space that is otherwise available. 
From the $m_{\tilde{\mu}_{R}}$ - $m_{\tilde{\chi}^{0}_{1}}$ 
plane we see that the upper bound on $m_{\tilde{\mu}_{R}}$ is $\sim$ 550 GeV.
Similarly, in $m_{\tilde{\nu}_{\mu}}$ - $m_{\tilde{\chi}^{0}_{1}}$ plane, we see that $g_{\mu}-2$ anomaly constraint puts an upper bound on the $m_{\tilde{\nu}_{\mu}}$ to $\sim700$ GeV.

We see in the the $m_{\tilde{\tau}_{\text{\tiny L}}}$ - $m_{\tilde{\chi}^{0}_{1}}$ plane that the sparticles and Higgs mass constraints put a lower bound of 900 GeV on $m_{\tilde{\tau}_{\text{\tiny L}}}$ which is in way above the experimental limit of 41 GeV. We can also see that the $g_{\mu}-2$ anomaly constraint applies a lower bound on $m_{\tilde{\tau}_{\text{\tiny L}}}$ of 1500 GeV while the upper bound on $m_{\tilde{t_{\text{\tiny L}}}}$ from this constraint is $\sim 2500$ GeV.

As for the $m_{\tilde{t}_{\text{\tiny L}}}$ - $m_{\tilde{\chi}^{0}_{1}}$ plane, we can see that the sparticles and Higgs mass bounds put a lower bound of 1350 GeV on left-handed top squark mass which is way higher then experimental lower bound of 800 GeV. We can also see that the lower bound on $m_{\tilde{t}_{\text{\tiny L}}}$  is 1800 GeV, from $g_{\mu}-2$ anomaly constraint while the upper bound is 2600 GeV. In these third family sleptonic planes, we see that the $g_{\mu}-2$ constraint allowed region is super-imposed over the B Physics allowed region for heavier sleptons. So we can have heavy third family sfermions, which are required for the Higgs mass correction, and are compatible with $g_{\mu}-2$ and all other collider bounds. 
 \begin{figure}[H]
 	\captionsetup{width=0.90\textwidth}
 	\begin{subfigure}{.5\textwidth}
 		\includegraphics[width=8cm,height=6cm]{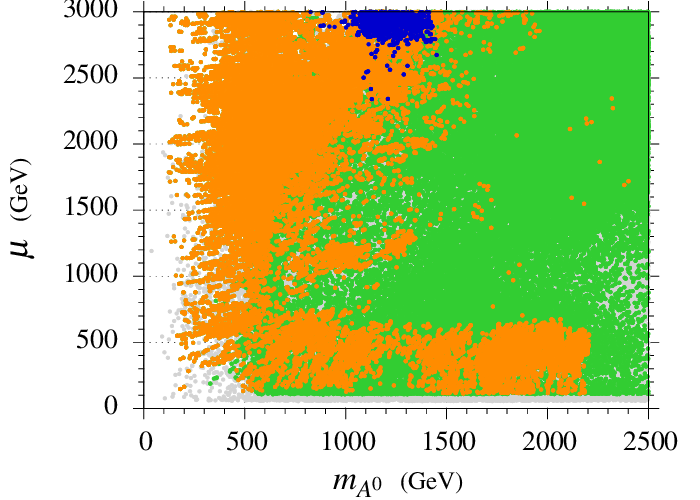}
 		\label{fig:sfig52}
 	\end{subfigure}
 	\begin{subfigure}{.5\textwidth}
 		\includegraphics[width=8cm,height=6cm]{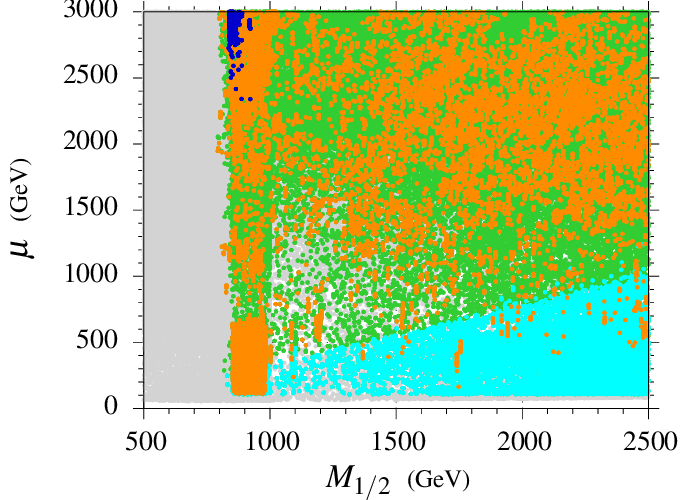}
 		\label{fig:sfig53}
 	\end{subfigure}
 	\caption{Plots in the ${\mu}$ - $m_{A^{0}}$ and$\mu$ - $M_{1/2}$ planes, color coding is the same as in Fig.~\ref{fig:fig1}. In $\mu$ - $M_{1/2}$ plane the cyan colored points prefer the Higgsino like dark matter.}
 	\label{fig:fig5}
 \end{figure}
 In Fig. {\ref{fig:fig5}}, we present our results in  ${\mu}$ - $m_{A^{0}}$ and $\mu$-$M_{1/2}$ planes where $\mu$ is the bilinear Higgs mixing parameter and $m_{A^{0}}$  is mass of the CP-odd Higgs particle. We can see that the lower mass bound, from the $g_{\mu}-2$ constraint, on $m_{A^{0}}$ is $\sim 650$ GeV and upper limit is $\sim1500$ GeV  whereas the lower bound on the $\mu$ parameter is $\sim 2300$ GeV. It can be seen that the B Physics constraints prefer the lower mass of $m_{A^0}$ as the MSSM Wilson coefficients (mentioned in the section of effective theory of B decays) inversely depends on $m_{A^0}^2$.  
 
 In the $\mu$ - $M_{1/2}$ plane, the cyan color is for the points where neutralino is higgsino like. So we can see that, in sMSSM, the $g_{\mu} -2$ constraint does not favor higgsino-like dark matter.
 \begin{figure}[H]
 	\captionsetup{width=0.90\textwidth}
 	\begin{subfigure}{.5\textwidth}
 		\includegraphics[width=8cm,height=6cm]{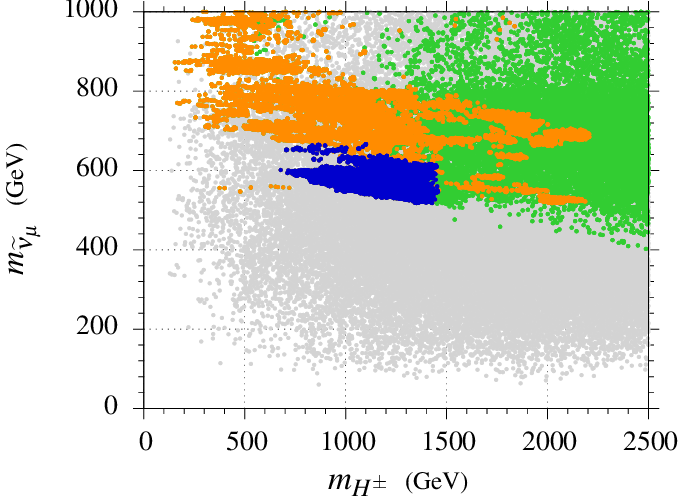}
 		\label{fig:sfig62}
 	\end{subfigure}
 	\begin{subfigure}{.5\textwidth}
 		\includegraphics[width=8cm,height=6cm]{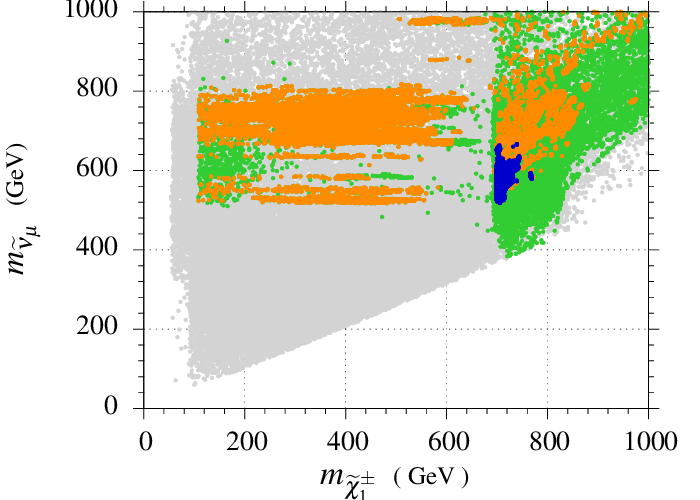}
 		\label{fig:sfig63}
 	\end{subfigure}
 	\caption{Plots in the $\tilde{\nu}_{\mu}$ - $m_{H^{\pm}}$ and
 		$\tilde{\nu}_{\mu}$ - $m_{\tilde{\chi}^{\pm}_{1}}$ planes, color coding is the same as in Fig.~\ref{fig:fig1}. In $\mu$ - $M_{1/2}$ plane the cyan colored points prefer the Higgsino like dark matter.}
 	\label{fig:fig6}
 \end{figure}
 In Fig. {\ref{fig:fig6}} we show our plots in $m_{\tilde{\nu}_{\mu}}$- $m_{H^{\pm}}$ and $m_{\tilde{\nu}_{\mu}}$- $m_{\tilde{\chi}^{\pm}}$ planes. In  $m_{\tilde{\nu}_{\mu}}$ - $m_{H^{\pm}}$ plane, we can see the lower limit on $m_{H^{\pm}}$ is $\sim650$ GeV and the allowed mass ranges for $m_{\tilde{\nu}_{\mu}}$ decrease with increasing $m_{H^{\pm}}$. So in sMSSM we can have a lighter $\widetilde{\nu}_{\mu}$ and a heavy $H^{\pm}$ where the first one is required for MSSM correction to $g_{\mu}-2$, while the second one is required for satisfying constraints from $B_{d} \to K^{*} \mu^{+} \mu^{-}$.

  In $m_{\tilde{\nu}_{\mu}}$- $m_{\tilde{\chi}^{\pm}}$ plane, we can see that the lower and upper mass bounds on chargino, $\tilde{\chi}^{\pm}$, due to $g_{\mu}-2$ constraint are $\sim650$ GeV and $\sim750$ GeV, respectively. The inverse square dependence of MSSM contributions to B Physics observable on charginos mass and charged Higgs mass, can be seen in Fig. {\ref{fig:fig6}. It can also be seen that for $m_{\tilde{\chi}^{\pm}}$ $ < $ 450 GeV, a lower mass bound of $\sim 500$ GeV is applied on $m_{\widetilde{\nu}_{\mu}}$ from the sparticles and Higgs mass constraints.

  \begin{figure}[H]
  	\captionsetup{width=0.90\textwidth}
  	\begin{subfigure}{.5\textwidth}
  		\centering
  		\includegraphics[width=8cm,height=6cm]{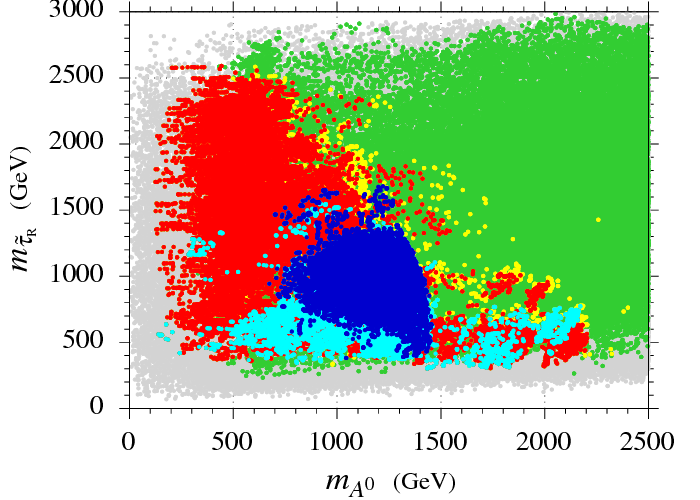}
  		\label{fig:sfig102}
  	\end{subfigure}
  	\begin{subfigure}{.5\textwidth}
  		\centering
  		\includegraphics[width=8cm,height=6cm]{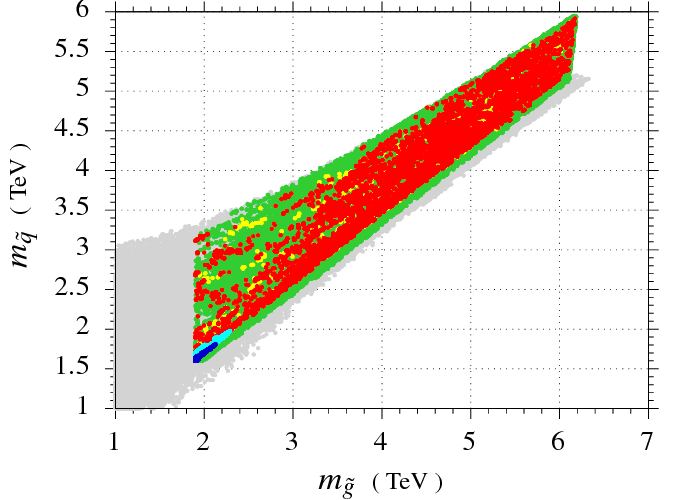}
  		\label{fig:sfig104}
  	\end{subfigure}
  	
  	\caption{Plots in the  $ m_{\tilde{\tau_{\text{\tiny R}}}}$ - $m_{{A}^{0}}$ and $m_{\tilde{q}}$ - $m_{\tilde{g}}$ planes. Yellow points corresponds to points with 95 $\%$ CL and red are with 68 $\%$ CL for $\chi^2$ of B Physics observables given in equation \eqref{chiSquare}. The cyan color points from a subset of yellow points and satisfy the $g_{\mu}-2$ anomaly constraints in 2-$\sigma$ range. The rest of the colors are the same as in Fig. \ref{fig:fig1}. Also yellow points are a subset of green points.}
  	\label{fig:fig10}
  \end{figure}
 In Fig. \ref{fig:fig10}, we present plots for $ m_{\tilde{\tau_{\text{\tiny R}}}}$ - $m_{{A}^{0}}$ and $m_{\tilde{q}}$ - $m_{\tilde{g}}$ planes, where $ m_{\tilde{\tau_{\text{\tiny R}}}}$ is the mass of the SUSY partner of right-handed tauon. Color coding for grey, green and blue points is the same as in Fig. \ref{fig:fig1}. Yellow and red points, subset of green points, have 95 $\%$ CL and 68 $\%$ CL for $\chi^2$ of B Physics observables given in equation \eqref{chiSquare}, respectively. The cyan color points from a subset of yellow points and satisfy the $g_{\mu}-2$ anomaly constraints at 2-$\sigma$ range.
 
In $ m_{\tilde{\tau_{\text{\tiny R}}}}$ - $m_{{A}^{0}}$ plane of Fig. \ref{fig:fig10}, we can see that the for heavier $m_{{A}^{0}}$, B Physics constraints prefer lighter $ m_{\tilde{\tau_{\text{\tiny R}}}}$. The $g_{\mu}-2$ anomaly 2-$\sigma$ constraint bounds on $m_{{A}^{0}}$, are almost identical to its B Physics mass bounds. The lower mass bound on $ m_{\tilde{\tau_{\text{\tiny R}}}}$ is $\sim$300 GeV which is also consistent with lower mass bound applied by B Physics constraints. $g_{\mu}-2$ anomaly constraint 1-$\sigma$ and 2-$\sigma$ bounds put an upper mass bound of 1700 GeV on $\tilde{\tau_{\text{\tiny R}}}$ mass.

In $m_{\tilde{q}}$ - $m_{\tilde{g}}$ plane, we can see that the B Physics constraints shrink the allowed regions from sparticles and Higgs mass constraints. Also we can see that the upper mass bound on gluino from 2-$\sigma$ range of $g_{\mu}-2$ anomaly is $\sim$ 2300 GeV which is 200 GeV above the 1-$\sigma$ allowed range. There is also 200 GeV increase in the upper bound of squarks mass, $m_{\tilde{q}}$ if we take 2-$\sigma$ range of $g_{\mu}-2$ anomaly constraint.

The lower and upper mass bounds on $m_{\tilde{\tau}_{\text{\tiny L}}}$, $m_{\tilde{\mu}_{\text{\tiny R}}}$, $m_{\tilde{\tau}_{\text{\tiny L}}}$, $m_{\tilde{t}_{\text{\tiny L}}}$, $m_{\tilde{\tau}_{\text{\tiny L}}}$ due to $g_{\mu}-2$ anomaly 2-$\sigma$ constraint are  100 GeV and 450 GeV, 350 GeV and 800 GeV, 500 GeV and 800 GeV, 1350 and GeV, 940 and 2450 GeV, respectively.

The total number of data points collected for our scan is more than a million. After successive application of constraints the number of data points decreases. We have $\sim$0.45 million green points, $\sim$0.16 million yellow points and $\sim$0.125 million red points so there is a difference of nearly thirty thousand points between yellow(points with 95$\%$ CL) and red(points with 68$\%$ CL) ones. 
 \begin{figure}[t]
 	\captionsetup{width=0.90\textwidth}
 	\begin{subfigure}{.5\textwidth}
 		\centering
 		\includegraphics[width=8cm,height=6cm]{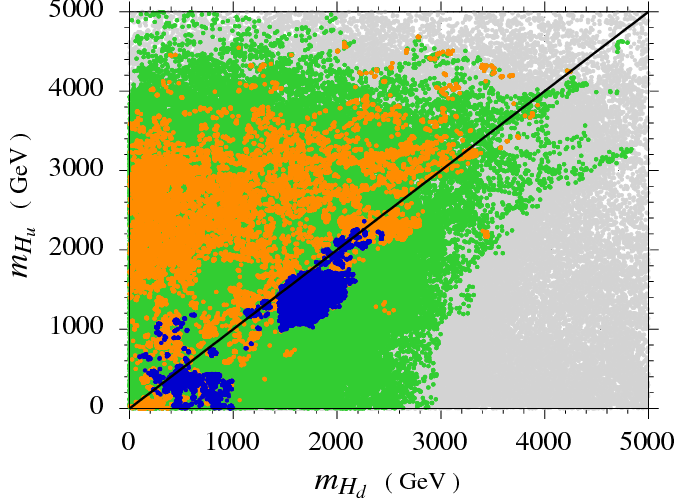}
 		\label{fig:sfig71}
 	\end{subfigure}
 	\begin{subfigure}{.5\textwidth}
 		\centering
 		\includegraphics[width=8cm,height=6cm]{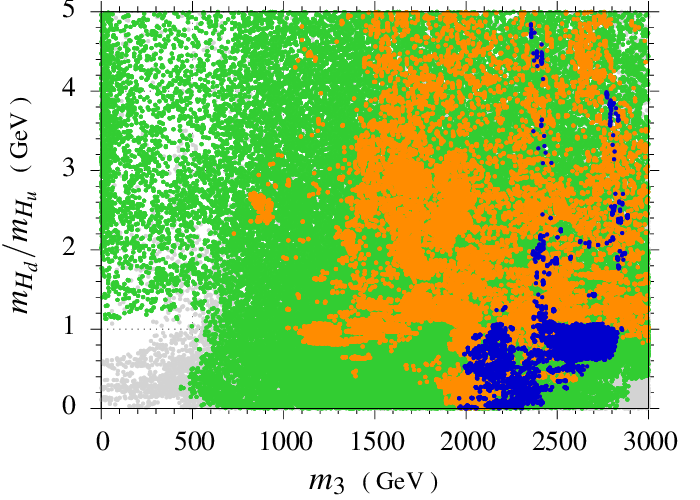}
 		\label{fig:sfig72}
 	\end{subfigure}
 	\caption{Plots in the  $m_{H_{u}}$ - $m_{H_{d}}$ and ${m_{H_{u}}}/{m_{H_{d}}}$ - $m_3$ planes, color coding is the same as in Fig.~\ref{fig:fig1}.}
 	\label{fig:fig7}
 \end{figure}

 In Fig.~\ref{fig:fig7},  we present plots for $m_{H_{u}}$ - $m_{H_{d}}$ and $m_{H_{u}/m_{H_{d}}}$- $m_{3}$ planes. In $m_{H_{u}}$ - $m_{H_{d}}$ plane, we can see that the upper limits on $m_{H_{u}}$ and $m_{H_{d}}$, from $g_{\mu} -2 $ constraint, are 2400 GeV and 2500 GeV, respectively. It can also be seen in Fig.~{\ref{fig:fig7}} that for $m_{H_{d}}/m_{H_{u}}$=$1$, we get data consistent with both the $g_{\mu}-2$ and B Physics constraints. So we next do a systematic analysis of the NUHM-I based sMSSM model in which $m_{H_{u}}=m_{H_{d}}$ but we still retain the splitting between the $1^{\text{st}}/2^{\text{nd}}$ and $3^{\text{rd}}$ generation scalars.
\subsection{sMSSM with NUHM-I}
In the NUHM-I model, we have one parameter($m_{10}$) to replace both $m_{H_{u}}$ and $m_{H_{d}}$.
One would expect this model to be slightly more constrained than the one discussed previously.
 \begin{figure}[H]
 	\captionsetup{width=0.90\textwidth}
 	\begin{subfigure}{.5\textwidth}
 		\centering
 		\includegraphics[width=8cm,height=6cm]{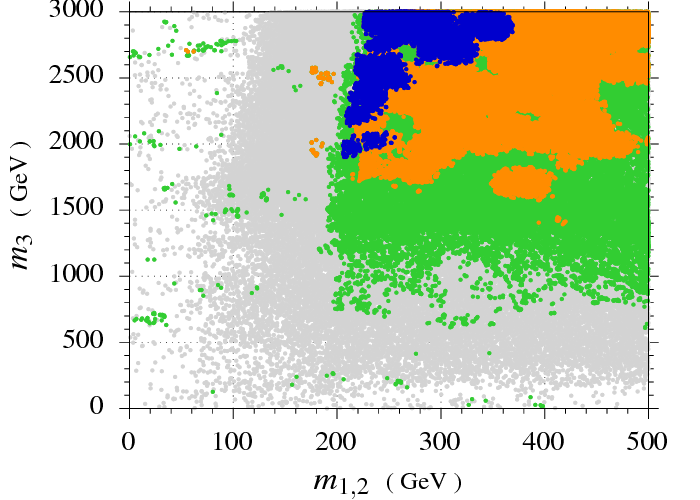}
 		\label{fig:sfi81}
 	\end{subfigure}
 	\begin{subfigure}{.5\textwidth}
 		\centering
 		\includegraphics[width=8cm,height=6cm]{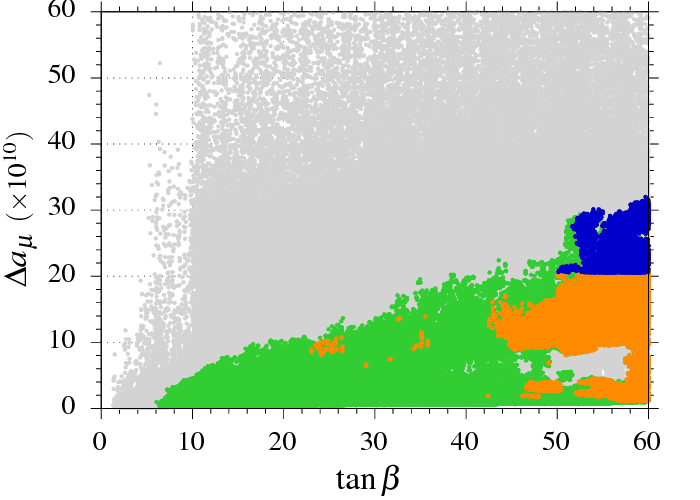}
 		\label{fig:sfig82}
 	\end{subfigure}
 	 	\begin{subfigure}{.5\textwidth}
 	 		\centering
 	 		\includegraphics[width=7cm,height=6cm]{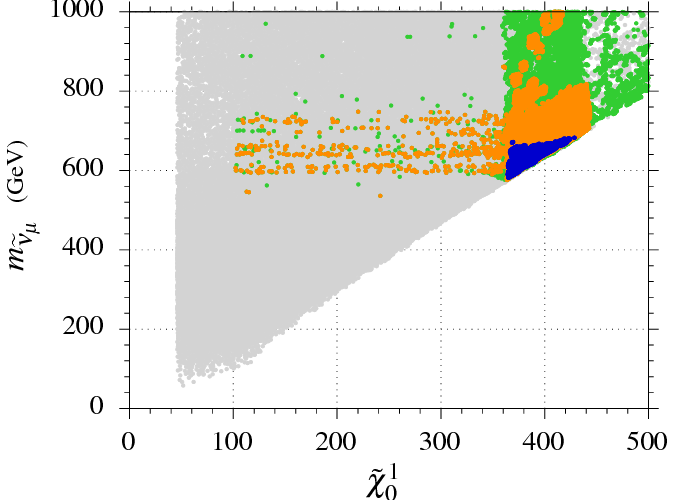}
 	 		\label{fig:sfig83}
 	 	\end{subfigure}
 	 	\begin{subfigure}{.5\textwidth}
 	 		\centering
 	 		\includegraphics[width=7cm,height=6cm]{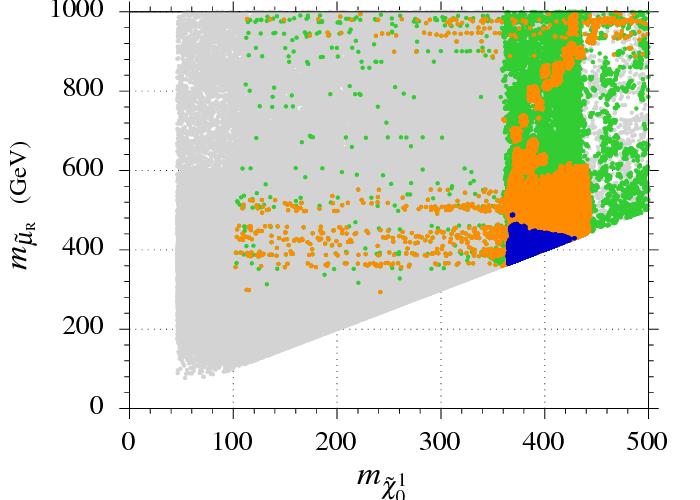}
 	 		\label{fig:sfig84}
 	\end{subfigure}
 	\caption{Plots in the  $\Delta a_{\mu}$ - $m_{\tilde{\mu}_{R}}$ and planes, color coding is the same as in Fig.~\ref{fig:fig1}.}
 	\label{fig:fig8}
 \end{figure}
 In Fig.~\ref{fig:fig8}, we show comparison of the $1^{\text{st}}$ two families sfermions masses verses the third family sfermions mass. The lower mass bound on $m_{1,2}$ is 200 GeV but the upper bound is 360 GeV which is same as for NUHM-II case. The allowed range for $m_{3}$ has same upper and lower bounds as those of NUHM-II. The allowed region boundaries are more contracted as compared to NUHM-II case as they form a subset of the allowed region of NUHM-II allowed region.
 
 In $\Delta a_{\mu}$ - $\tan \beta$ plane we can see that the $\tan\beta$ $\lesssim$ 6 is not allowed from sparticles mass bounds. B Physics constraints don not allow $\tan\beta$ $\lesssim$ 22. $g_{\mu}-2$ anomaly constraint put a lower bound on $\tan\beta$ at $\tan\beta$ $\sim$ 50. All these ranges are more strict as compared to NUHM-II ranges.

In $m_{\tilde{\mu}_{\text{\tiny R}}}$-$m_{\tilde{\chi_{0}^{1}}}$ plane, we can see that the upper bounds on $m_{\tilde{\mu}_{\text{\tiny R}}}$ is 500 GeV and lower mass bound is 360 GeV. We can also see that this plane also explains the smuon-neutralino coannihilaion in NUHM-I type sMSSM. 

In $m_{\tilde{\nu}_{\mu}}$-$m_{\tilde{\chi_{0}^{1}}}$ plane we can see that the allowed mass range for smuon-neutralino, from $g_{\mu}-2$ anomaly, is $580$ GeV $\lesssim$ $m_{\tilde{\nu}_{\mu}}$ $\lesssim$ 680 GeV.
 
\begin{table}[H]
\captionsetup{width=0.95\textwidth}
	\footnotesize
	
	\centering
	\begin{tabular}{|ccccccc|cccc|c|}
		\hline
	{$m_{1,2}$} & {$m_{3}$} & {$M_{1/2}$} & {$A_{0}$} & $\tan\beta$ &
		$m_{H_{d}}$ & $m_{H_{u}}$ & $m_{\tilde{\mu}_{R}}$ & $m_{\tilde{\chi}^{0}_{1}}$ & $m_{\tilde{\nu}_{\mu}}$ & $m_{\tilde{g}}$ & $\Omega h^{2}$ \\
	\hline
	\hline
	{140} & {2355} & {876} & {-4470} & {57} &
	{786} & {1143} & {383} & {382} & {576} & {1975} & {0.1}\\
	{232} & {3004} & {846} & {-4940} & {57} &
	{1382} & {1382}  & {373}& {370}  & {590} & {1927} & {0.1}\\
	\hline
\end{tabular}
\caption{Demonstration of the possibility of getting the correct dark matter relic density for co-annihilation scenarios, ${\tilde{\mu}_{\text{\tiny R}}}$ co-annihilation for NUHM-II inspired sMSSM ($1^{st}$ row) and NUHM-I inspired sMSSM ($2^{nd}$ row). All masses are in GeV.}
\label{t1}
\end{table}
\section{Conclusion}
\label{conclusion}
We have investigated the sparticle spectrum of the flavor symmetry based Minimal Supersymmetric 
Standard Model (sMSSM) which is distinguished as having the third generation 
scalar mass parameter different from the first two. This splitting allows us to simultaneously 
satisfy constraints from B Physics and direct mass bounds that require a heavy SUSY spectrum and 
the muon $g_\mu-2$ constraint that favors a lighter SUSY spectrum. We have provided the allowed mass ranges which are compatible with the constraints from
branching ratios of radiative and pure leptonic decays $b \to s \gamma$,
$B_s \to \mu^{+} \mu^{-}$ and $B_{u} \to \tau \nu_{\tau}$, respectively, as well as
the sparticle and gluino mass constraints, the Higgs mass constraint and
the constraints from $B_{d} \to K^{*} \mu^{+}\mu^{-}$. In our analysis, we have 
used the latest experimental data available from the LHCb, the Muon $ g-2$ 
Experiment and the ATLAS and CMS collaborations.

We show that in the sMSSM with both universal and non-universal Higgs mass parameters ($m_{H_{u,d}}$) we 
are able to satisfy all the known experimental bounds and explain the Standard Model $g_\mu-2$ anomaly. 
We also show that this model can provide a dark matter candidate, which may belong to the smuon co-annihilation region in both universal and non-universal Higgs 
scenario.

 We have seen that the allowed range of $m_{\tilde{\chi}^{0}_{1}}$ is from 360 GeV to 420 GeV. We, furthermore, show that the upper limits on the stop and gluino masses are 1800 GeV and 2100 GeV, respectively.

\section*{Acknowledgments}
We thank Ilia Gogoladze, Ali Paracha, Cem Salih Un and Mehar Ali Malik for useful discussions. We acknowledge the use of super-computing facilities at Research Centre for Modeling and Simulations (RCMS) at National University of Sciences and Technology (NUST) and National Centre for Physics, Shahdra Velly Road, Islamabad, Pakistan for the work carried out and thank the same. 

\end{document}